\documentclass[10pt,twocolumn]{article}

\usepackage{leading}
\usepackage[margin=1in]{geometry}

\usepackage{times}
\usepackage{listings, textcomp}
\usepackage[T1]{fontenc}

\PassOptionsToPackage{hyphens}{url}\usepackage[colorlinks=true,urlcolor=blue,citecolor=blue,linkcolor=blue,plainpages=false]{hyperref}
\usepackage[normalem, normalbf]{ulem} 
\usepackage{flushend} 
\usepackage{xcolor}
\usepackage{multirow}
\usepackage{booktabs}
\usepackage{graphicx}
\usepackage{enumitem}
\usepackage{changes}
\usepackage{amsmath}
\usepackage[TABBOTCAP]{subfigure}
\begin{document}

\leading{12pt}

\title{\bf Scalable, Fast Cloud Computing with Execution Templates}
\author{Omid Mashayekhi, Hang Qu, Chinmayee Shah, Philip Levis\\
        Stanford University\\
        \{omidm,~quhang,~chshah\}@stanford.edu~~pal@cs.stanford.edu}

\date{}
\maketitle
\thispagestyle{empty}

\begin{abstract}
Large scale cloud data analytics applications are often CPU
bound. Most of these cycles are wasted: benchmarks written in C++ run
10-51 times faster than frameworks such as Naiad and Spark.  However,
calling faster implementations from those frameworks only sees
moderate (3-5x) speedups because their control planes cannot schedule
work fast enough.

This paper presents execution templates, a control plane abstraction
for CPU-bound cloud applications, such as machine learning.  Execution
templates leverage highly repetitive control flow to cache scheduling
decisions as {\it templates}.  Rather than reschedule hundreds of
thousands of tasks on every loop execution, nodes instantiate these
templates.  A controller's template specifies the execution across all
worker nodes, which it partitions into per-worker templates. To ensure
that templates execute correctly, controllers dynamically patch
templates to match program control flow.  We have implemented
execution templates in Nimbus, a C++ cloud computing framework.
Running in Nimbus, analytics benchmarks can run 16-43 times faster
than in Naiad and Spark. Nimbus's control plane can scale out to run
these faster benchmarks on up to 100 nodes (800 cores).

\end{abstract}

\section{Introduction}


The CPU has become the new bottleneck for analytics benchmarks and
applications. One recent study found that the big data benchmark
(BDBench), TCP decision support benchmark (TCP-DS), and production
workloads from Databricks were all CPU-bound. Improving network I/O
would reduce their median completion time by at most 2\% and improving
disk I/O would reduce their median completion time by at most
19\%~\cite{ousterhout15}.


At the same time, systems such as DMLL~\cite{dmll} and
DimmWitted~\cite{dimwitted} have shown it is possible to achieve
orders-of-magnitude improvements in CPU performance over frameworks
such as Spark~\cite{spark}.  Comparing the performance of C++ and
Spark implementations of two standard machine learning
benchmarks, we find that the C++ implementations run up to {\it 51
times faster}. Modern analytics frameworks are CPU-bound, but most of
these cycles are wasted.

One straw man solution to improve performance is to have a framework
call into C++ implementations of computational kernels, e.g., through
the Java Native Interface (JNI). In Section~\ref{sec:implications}, we
show that this only sees modest speedups (5x rather than 50x): worker
nodes spend 90\% of their cycles idle. The central Spark {\it
controller}, which is responsible for telling to workers to execute
tasks, cannot schedule tasks quickly enough. The framework's control
plane becomes a bottleneck and workers fall idle. In
Section~\ref{sec:evaluation} we show that Naiad~\cite{naiad}, another
framework, has similar control plane bottlenecks.

Current frameworks do not scale to run optimized tasks on many
nodes. They can either run on many nodes or run optimized tasks, but
not both, because the control plane cannot schedule tasks fast
enough. Prior scalable scheduling systems such as
Sparrow~\cite{sparrow}, Omega~\cite{omega}, Apollo~\cite{apollo},
Mercury~\cite{mercury}, Hawk~\cite{hawk} and Tarcil~\cite{tarcil} all
propose ways to distribute the scheduling of many jobs which together
overwhelm a single controller.  Scheduling a job requires
centralized state, and so for all these systems, tasks from a single
job still go through a single scheduler.  Optimized tasks, however,
mean that a {\it single} job can saturate a controller.

Section~\ref{sec:templates} presents {\it execution templates} a
control plane abstraction which scales to schedule optimized tasks on
many nodes.  The key insight behind execution templates is that
long-running CPU-bound computations are repetitive: they run the same
computation (e.g., a loop body) many times. Rather than reschedule
each repetition from scratch, a runtime caches scheduling decisions as
an execution template of tasks. A program invokes a template,
potentially creating thousands of tasks, with a single message. We
call this abstraction a template because it can cache some decisions
(e.g., dependencies) but fully instantiating it requires parameters
(e.g., task identifiers).



Section~\ref{sec:implementation} describes an implementation of
execution templates in Nimbus, a C++ analytics framework that
incorporates execution templates. Compared to Spark and Naiad,
benchmarks in Nimbus run 16-43 times faster. Rewriting benchmarks in
Spark and Naiad to use optimized tasks reduces their completion time
by a factor of 3.7-5. However, Section~\ref{sec:implications} shows
results that neither can scale out past 20 worker nodes
because the control plane becomes a bottleneck: running on more than
20 nodes {\it increases} completion time. Using execution templates,
implementations of these benchmarks in Nimbus scale out to 100 nodes
(800 cores), seeing nearly linear speedups.

Execution templates allow a centralized controller to handle tasks
shorter than 1ms, or 100 times shorter than what prior systems
support~\cite{sparrow}. This makes whole new applications possible. We
have ported PhysBAM, a graphical simulation library~\cite{physbam}
used in many feature films\footnote{PhysBAM is a cornerstone of
  special effects at Industrial Light and Magic and is also used
  Pixar.} to Nimbus.  PhysBAM has tasks as short as 100$\mu$s, yet
execution templates can execute extremely large simulations within 15\%
of the speed of PhysBAM's hand-tuned MPI libraries.

This paper makes five contributions:

\begin{enumerate}[leftmargin=*] 
  \setlength{\itemsep}{.5pt}

\item A detailed analysis of how Spark
  spends CPU cycles, finding that C++ implementations run 51 times faster and
  most of Spark's cycles are wasted due to runtime and programming
  language overheads (Section~\ref{sec:speedups}).

\item Results showing Spark and Naiad's control planes are a
  bottleneck when running optimized (C++) tasks and so they can only
  provide modest speedups (Section~\ref{sec:implications}).

\item Execution templates, a novel control plane abstraction that
  allows optimized tasks to run at scale
  (Section~\ref{sec:templates}).

\item The design of Nimbus, an analytics framework that incorporates
  execution templates and a data model based on mutable data objects
  which permit in-place modifications (Section~\ref{sec:implementation}).

\item An evaluation of execution templates, finding they allow Nimbus
  to run optimized tasks with almost no overhead, scaling out to 100
  nodes (800 cores) while running 30-43 times faster than Spark and
  16-23 times faster than Naiad. Execution templates also allow Nimbus to
  support large, complex applications with tasks as short as 100$\mu$s
  (Section~\ref{sec:evaluation}).

\end{enumerate}

Section~\ref{sec:implementation} provides details on the Nimbus
implementation of execution templates, including the dynamic program
analysis that ensures they execute properly despite variations in
control and data flow. Section~\ref{sec:related} presents related
work and Section~\ref{sec:conclusion} concludes.


\section{Motivation}

A recent study found that Spark analytics applications are
CPU-bound~\cite{ousterhout15}. Increasing server RAM and easy
parallelization means that many applications can keep their entire
working set in RAM and completion time is limited by CPU performance.

This section motivates the need for a new control plane in cloud data
analytics frameworks. It starts by examining where Spark's CPU cycles
go: 98\% of them are wasted. Re-implementations in C++ run up to 51 times
faster.  However, if a Spark job uses these faster re-implementations,
it only sees modest (5x) speedups because the control plane
(messages to schedule and dispatch tasks)
become the bottleneck. The section concludes by observing an important property
of CPU-bound applications, that their control flow and execution
exhibits very regular patterns, which can be calculated, cached and
reused.

\subsection{Where the Cycles Go}\label{sec:speedups}

\begin{figure}
\centering
\includegraphics[width=3.0in]{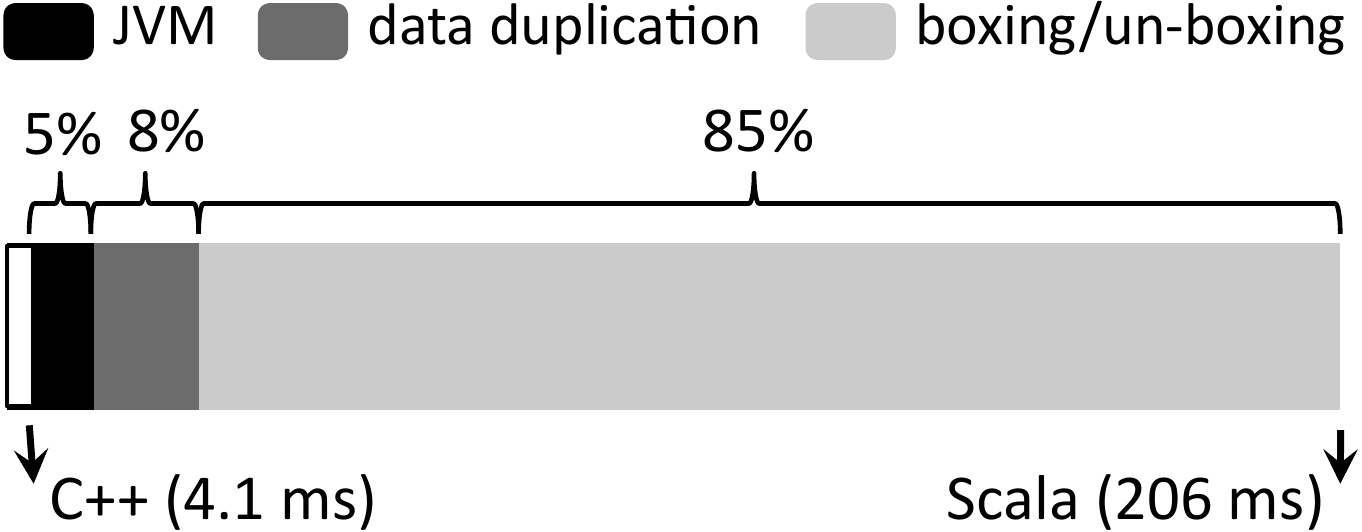}
\caption{Logistic regression execution time implemented in Scala and C++. C++
is 51 times faster than Scala. These results are averaged over 30
iterations and discard the first iteration to allow Java Virtual Machine (JVM)
to warm up and just-in-time compile.}
\label{fig:gradient}
\end{figure}

Frameworks such as Spark~\cite{spark} and Naiad~\cite{naiad} focus on
applications whose data sets can fit in memory when spread across
many nodes.  At the same time, a push for greater programmer
productivity has led them to support higher-level languages: 70\% of
Spark applications are written in Scala~\cite{nvl-platformlab}.

\begin{table}
\centering
\setlength\tabcolsep{5pt} 
\begin{tabular}{lrrr} \toprule[2pt]
{\bf Code}  & {\bf Nodes} & {\bf Task Length} & {\bf Completion Time} \\ \midrule[1pt]
{\bf Scala} & 100 nodes   & 206ms             &  2.86s                \\ 
{\bf C++}   & 100 nodes   &   4ms             &  1.00s                \\ 
{\bf C++}   & 20 nodes    &  20ms             &  0.53s                \\ \bottomrule[2pt]

\end{tabular}
\caption{Effect of running optimized logistic regression tasks
  in Spark. Although C++ tasks can run 51 times faster, a job using
  C++ tasks on 100 nodes only runs 2.8x faster. It  
  run 5x faster when run on 20 nodes. Both are much slower than the
  expected speedups and 20 nodes is faster than 100 due to the control
  plane being unable to schedule tasks fast enough.}
\label{tab:melting}
\end{table}

These two trends (in-memory datasets and higher-level languages)
conflict: for applications that operate on in-memory data, higher-level
language overheads become significant.  Figure~\ref{fig:gradient}
shows the execution time of logistic regression, a common analytics
benchmark, implemented in Spark using Scala and implemented in C++.
The C++ implementation runs {\it 51 times faster} than the Spark one.

This poor performance has three major causes.\footnote{To determine the cause
of this slowdown, we configured the JVM to output the JIT assembly and
inspected it. We inserted performance counters in the Scala code re-inspected
the assembly to verify they captured the correct operations.  To separate the
cost of Scala from JVM bytecode interpretation, we decompiled the JVM bytecodes
Scala generated into Java, rewrote this code to remove its overheads,
recompiled it, and verified that the bytecodes for the computational operations
remained unchanged.} First, since Scala's generic methods cannot use primitive
types (e.g., they must use the {\tt Double} class rather than a {\tt double}),
every generic method call allocates a new object for the value, boxes the value
in it, un-boxes for the operation, and deallocates the object. In addition to
cost of a {\tt malloc} and {\tt free}, this results in millions of tiny objects
for the garbage collector to process. 85\% of logistic regression's CPU cycles
are spent boxing/un-boxing.

Second, Spark's resilient distributed datasets (RDDs) forces methods
to allocate new arrays, write into them, and discard the source array.
For example, a {\tt map} method that increments a field in a dataset
cannot perform the increment in-place and must instead create a whole
new dataset.  This data duplication adds an additional factor of
$\approx2$x slowdown.


Third, using the
Java Virtual Machine has an additional factor of $\approx3$x
slowdown over C++.  This result is in line with prior studies, which have
reported 1.9x-3.7x for computationally dense
codes~\cite{google-loops,gheradi3d}. In total, this results in Spark
code running 51 times slower than C++.

\subsection{Implications of Optimized Tasks}\label{sec:implications}


To determine how much tasks running at C++ speeds could improve
performance, we replaced the logistic regression benchmark's Spark
Scala code with loops that take as long as the C++ implementations.
This represents the best-case performance of Spark calling into a
native method (there is no overhead).

Table~\ref{tab:melting} shows the results. While the computational
tasks run 51 times faster, on 100 nodes the overall computation
only runs 2.8 times faster. Worker nodes spend most of the time idle
because the central Spark controller
cannot schedule tasks fast enough.  Each core can execute 250 tasks
per second (each task is 4ms), and 100 nodes (800 cores) can execute
200,000 tasks per second. We measured Spark's controller to be able to
issue $\approx$8,000 tasks per second.

This control plane bottleneck is not unique to Spark.
Naiad~\cite{naiad} is the best available \textit{distributed} cloud
framework. In Naiad, worker nodes directly coordinate with one another
rather than acting through a central controller. While Naiad code is
in C\# rather than Scala and so sees overall better performance than
Spark, its all-to-all synchronization also becomes a bottleneck above
20 nodes. We defer detailed experimental results on Naiad to
Section~\ref{sec:data-analytics}.

Scheduling techniques such as Sparrow~\cite{sparrow},
Omega~\cite{omega}, Apollo~\cite{apollo}, Mercury~\cite{mercury}, Hawk~\cite{hawk} and
Tarcil~\cite{tarcil}, address the scheduling bottleneck that occurs
when there are many concurrent jobs. In aggregate,
many jobs  can execute more tasks per second than a single
controller can schedule. But since these jobs share underlying
computing resources, they need to be scheduled cooperatively to prevent
overloading workers or contention. Each of these systems propose ways
for many separate,  per-job controllers to coordinate their
resource allocation and scheduling decisions. These systems all solve
the problem of when the aggregate task rate of {\it many} jobs is
greater than what one controller can handle.  Optimized tasks,
however, mean that {\it single} job can saturate a controller.
None of these systems can distribute a single job's scheduling.

\subsection{Observation: Repetition}\label{sec:repetitive}

Cloud computing applications are increasingly advanced data analytics
including machine learning, graph processing, natural language
processing, speech/image recognition, and deep learning.  These
applications are usually implemented on top of frameworks such as
Spark~\cite{spark} or Naiad~\cite{naiad}, for seamless parallelization
and elastic scalability. A recent survey~\cite{spark-survey} of Spark
users, for example, shows 59\% of them use the Spark machine learning
library~\cite{spark-ml}.  Efforts such as Apache Mahout~\cite{mahout}
and Oryx~\cite{oryx} provide machine learning libraries on top of
Spark.  Cloud providers, in response to this need, now offer special
services for machine learning models~\cite{azure-ml, amazon-ml}.

One important property of analytics jobs is their computations have
repetitive patterns: they execute a loop (or set of nested loops)
until a convergence condition. The Ernest system~\cite{ernest}, for
example, leveraged this observation for predicting the performance
and managing resources. Logistic regression, for example, often
executes until parameters have converged and are no longer changing or
a large fixed number of iterations (whichever happens first).

\begin{figure}
    \centering
    \includegraphics[width=2.5in]{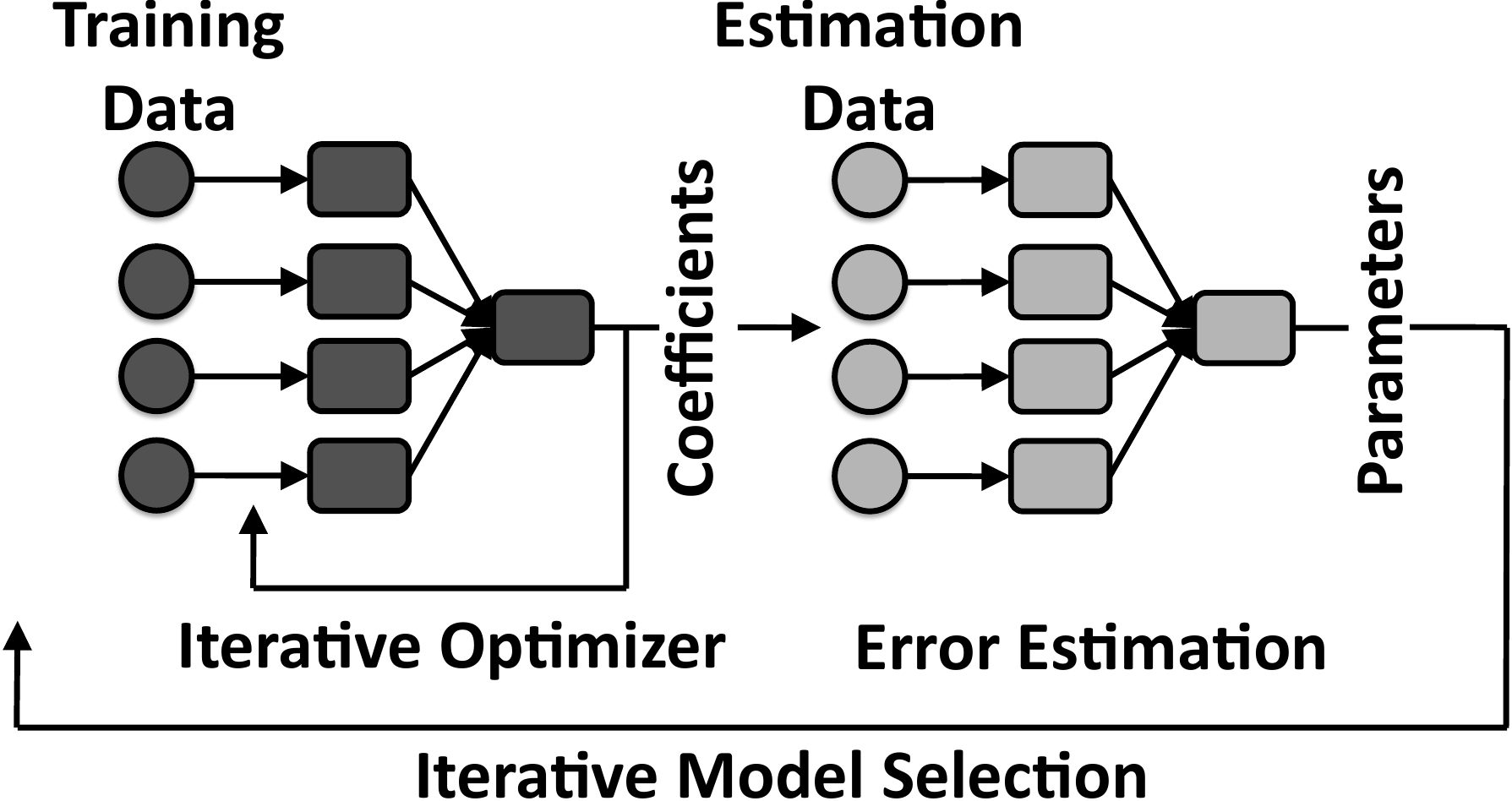}
    \caption{Execution graph of training a regression algorithm. It is
    iterative with an outer loop for updating model parameters based on the
    estimation error, and an inner loop for optimizing the feature
    coefficients.}
    \label{fig:cross-validation}
\end{figure}

For example, Figure~\ref{fig:cross-validation} shows the execution
graph of the hold-out cross validation method, a common machine
learning method used for training regression
algorithms~\cite{ml-book}. It has two stages, training and estimation,
which form a nested loop.  The training stage uses an iterative
algorithm, such as gradient descent, to tune coefficients. The
estimation stage calculates the error of the coefficients and feeds
this back into the next iteration of the training phase.

Each iteration generates the same tasks and schedules them to the same
nodes (those that have the data resident in memory).  Re-scheduling
each iteration repeats this work.  This suggests that a control plane
cached these decisions and reused would schedule tasks much
faster and scale to support fast tasks running on more nodes. The next
section describes execution templates, a control plane abstraction
that achieves this goal.

\begin{figure}
    \centering
    \includegraphics[width=2.5in]{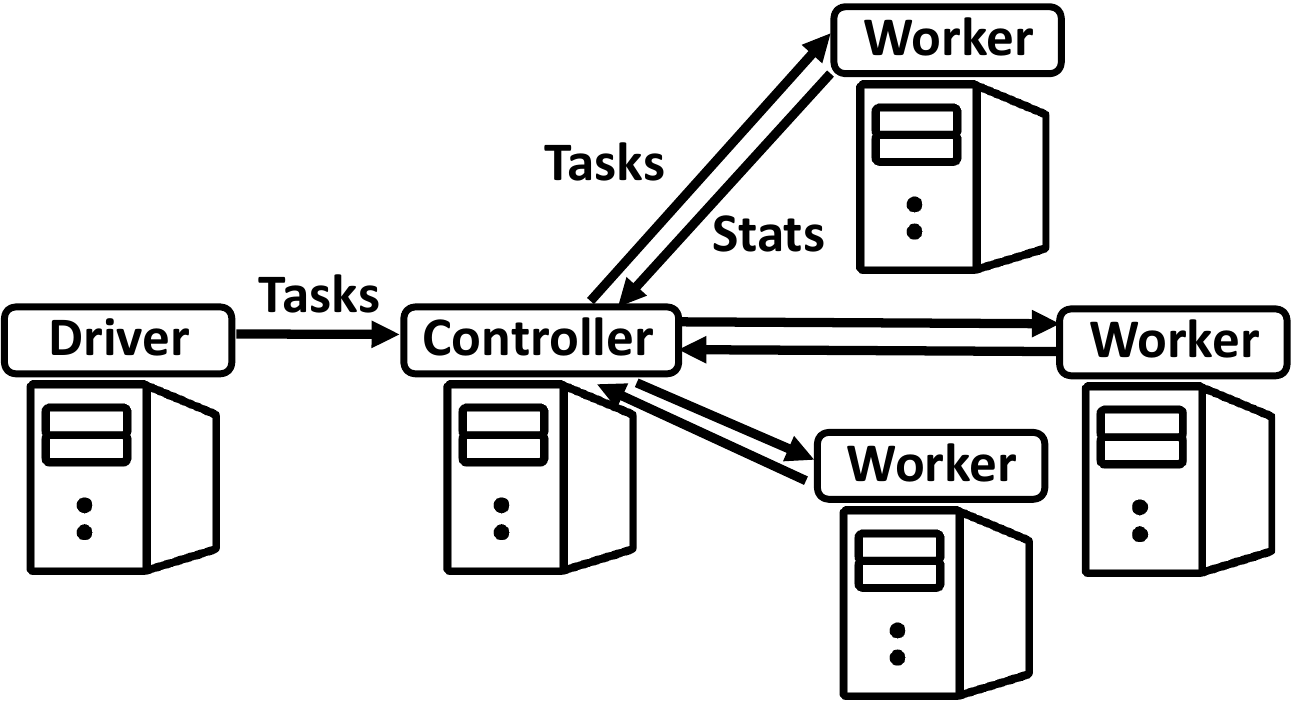}
    \caption{Architecture of a canonical cloud framework: a driver program
    specifies the application logic for a centralized controller, which drives
    the worker nodes to execute tasks.}
\label{fig:architecture}
\end{figure}

\begin{figure}
\lstnewenvironment{code}[1][]%
{
  \noindent\minipage{\linewidth}\medskip 
  \lstset{
    basicstyle=\ttfamily\footnotesize,
    showstringspaces=false,
    captionpos=b,
    breakatwhitespace=true,
    breaklines=false,
    tabsize=2,
    morekeywords={while, beginning, end, basic, block},
    morecomment=[s]{...}{...},
    upquote=true,
    escapeinside={@}{@},
    label={lst:api-code}
  }
}
{\endminipage}
\begin{code}
while (error > threshold_e) {
  while (gradient > threshold_g) { 
    // Start 1
    gradient = Gradient(tdata, coeff, param)
    coeff += gradient 
    // End 1
  } 
  // Start 2
  error = Estimate(edata, coeff, param) 
  param = update_model(param, error)    
  // End 2 
}
\end{code}
\caption{Driver program pseudocode for the iterative application in
  Figure~\ref{fig:cross-validation}. There are two basic blocks.
  {\tt Gradient} and {\tt Estimate} are both parallel
  operations that execute many tasks on partitions of data.}
\label{fig:driver-code}
\end{figure}

\section{Execution Templates}
\label{sec:templates}

We describe execution templates, a control plane abstraction for cloud
computing. Execution templates make it possible for workers to
inexpensively generate and execute large batches of tasks. 
If a program has a loop in it, rather than resend all of the tasks for that
loop to the controller on every iteration, it should instead send them
once. For subsequent iterations, it can tell the controller ``execute
those tasks again.'' The same is true for the controller; rather than
resend all of the tasks to the workers, it can tell each worker to
``execute those tasks again.'' 

The execution and control structure of cloud frameworks places
requirements on how templates operate.  Figure~\ref{fig:architecture}
shows the architecture of a cloud computing framework. A {\it driver}
program generates tasks, which it sends to a centralized {\it
  controller}. The driver and controller may or may not reside on the
same node. The controller processes these tasks and dispatches them to
a cluster of {\it workers}.  The controller balances load across
workers and recovers execution when one fails.

\begin{table}
\begin{tabular}{ll} \toprule[2pt]
{\bf Execution template} & {\bf JIT compiler}\\ \midrule[1pt]
Template                 & Function \\
Task (Driver$\rightarrow$Controller) & Bytecode instruction \\
Task (Controller$\rightarrow$Worker) & Native instruction \\
Data object              & Register \\
\bottomrule[2pt]
\end{tabular}
\caption{Execution templates are analogous to a just-in-time
compiler for a data analytics control plane.}
\label{tab:analogy}
\end{table}

Templates optimize repeated control decisions. In this way, they are
similar to a just-in-time (JIT) compiler for the control plane.  A JIT
compiler transforms blocks of bytecodes into native instructions;
execution templates transform blocks of tasks into dependency graphs
and other runtime scheduling structures.  Table~\ref{tab:analogy}
shows the correspondences in this an analogy: an execution template is
a function (the granularity JIT compilers typically operate on), a
task from the driver to the controller is a bytecode instruction, and
a task executing on the worker is a native instruction.

The rest of this section describes six requirements for how templates
operate.  While the analogy to JIT compilation fits well and many of
these requirements follow from it, the driver-controller-worker
execution model adds an additional requirement, the need to validate and patch
templates before executing them.

\medskip\noindent\textbf{1. Templates must be dynamically generated.}
Controllers and workers do not have the driver program. They receive a
stream of tasks, which they dynamically schedule and execute. They
therefore need to generate templates in response to this dynamic
stream of information. Furthermore, because a controller can
dynamically shift how it schedules tasks to workers (e.g., in response
to load imbalance or changing resources), it needs to be able to
correspondingly dynamically create new templates. Put another way,
templates cannot be statically compiled: they must instead be created
just-in-time.

\medskip\noindent\textbf{2. Templates must be parameterizable.}
Similarly to how a program must be able to pass parameters to
just-in-time compiled functions, a driver must be able to pass
parameters to execution templates. Analytics jobs involve many
repetitions of the same loop or computation, but the repetitions are
not identical. The cross-validation job in
Figure~\ref{fig:cross-validation}, for example, updates {\tt
  parameters}, which are then passed to the optimizer block. Each
instantiation of the optimizer block must fill in {\tt parameters} to
the {\tt find\_gradient} tasks. In addition to data parameters,
templates also require control parameters, such as which task
identifiers to use, to ensure that two workers do not use the same
globally unique identifier.

\medskip\noindent\textbf{3. Workers must locally resolve
  dependencies.} Large blocks of tasks often have data dependencies
between them. For example, the line {\tt coeff += gradient} in
Figure~\ref{fig:driver-code} cannot run until the previous line
computing {\tt gradient} completes. For a worker to be able to execute
the tasks for both lines of code locally, without coordinating with
the controller, it must know this dependency and correctly determine
when that line of code can run. This is similar to how a CPU uses data
flow to know when it can execute an instruction that depends on the
output of other instructions.

\medskip\noindent\textbf{4. Workers must directly exchange
  data.} Optimized tasks read and write in-memory data objects on
workers. Often, within a single template, the output of a task on one
worker is needed as the input for a task on another. As part of
executing the template, the two workers need to directly exchange this
data. This is similar to how two cores accessing the same memory need
to be able to to update each other's caches rather than always write
through to main memory.

\medskip\noindent\textbf{5. Controllers must be able to quickly validate
  and patch templates.} The driver-controller-worker execution model
adds additional complexities that JIT compilers do not need to handle.
Just as function calls assume that arguments are in certain registers
or stack positions, when a controller generates execution templates
for workers, it must assume certain preconditions on where data is
located. However, a driver can insert new tasks at any time, which
might violate these preconditions. For example, it might insert
instructions that increment a variable and store it in a different
register. When a controller instantiates a template, it must validate
whether a template's preconditions hold, and if not, insert tasks to
patch it. In the above example, the controller needs to detect the
variable is now in a new register and issue a move instruction to put
it back in the register the function call expects.

\medskip\noindent\textbf{6. Templates must be fast.} Finally, as the
overall goal of templates is to allow the control plane to support
optimized tasks at scale, the performance gains of instantiating them
must be greater than their cost to generate and instantiate.

Execution templates are tightly entwined with a framework's data model
and execution. The next section describes a concrete implementation of
them in the context of an analytics framework designed to
execute optimized tasks at scale.

\section{Implementation}
\label{sec:implementation}

\begin{figure*}
\centering
\setlength\tabcolsep{-3pt}
\begin{tabular}{cc}
\subfigure[Simple task graph example with three tasks and three data objects.
The data flow among tasks forms a DAG. For example, task C reads the updated
data objects 2, and 3 after execution of task A, and B.]
{
\includegraphics[width=3.0in]{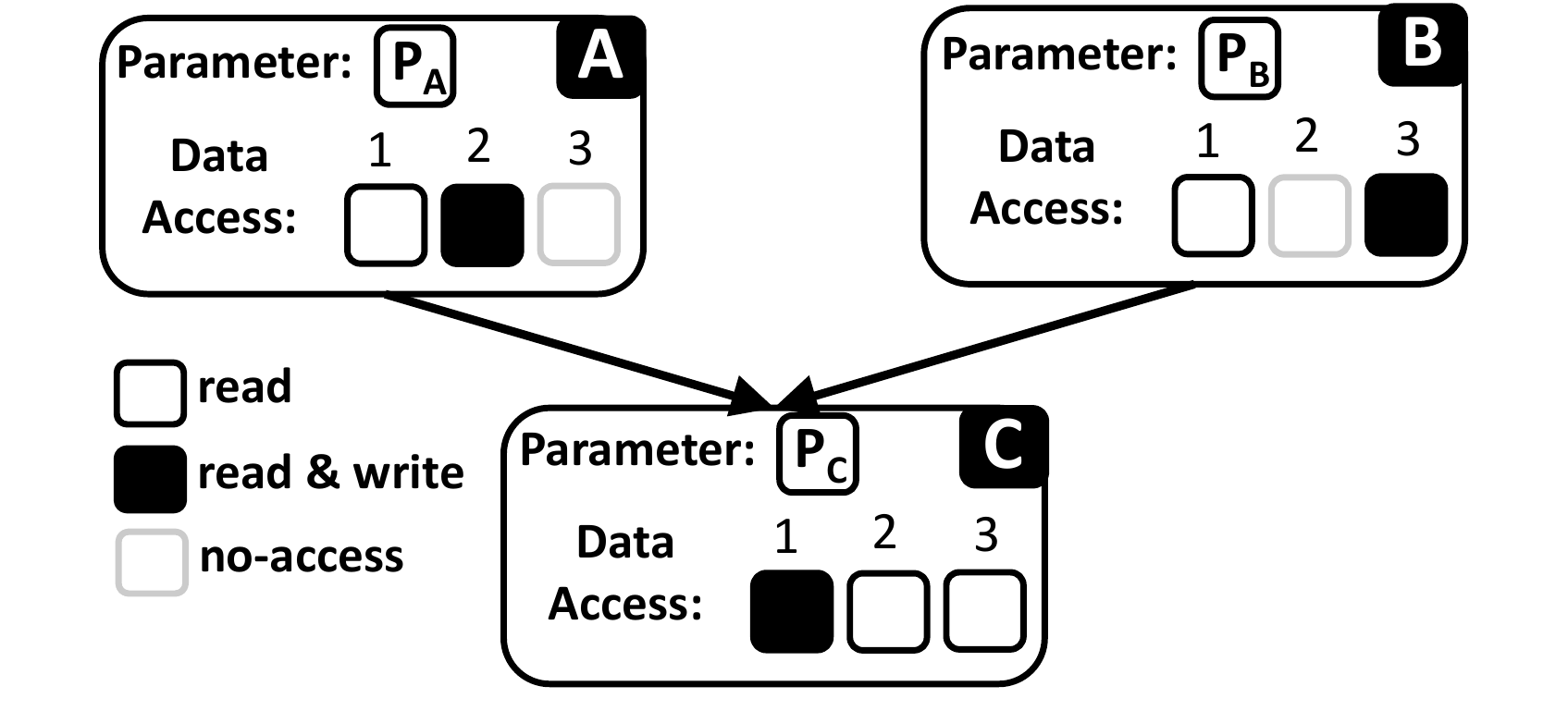}
\label{fig:simple-example}
}
&
\subfigure[ Mapping of the task graph in Figure~\ref{fig:simple-example} over
two workers. Each task graph embeds per worker task dependencies and data copy
among workers. Task graph dependencies allow workers to proceed without
controller's middling.]
{
\includegraphics[width=3.0in]{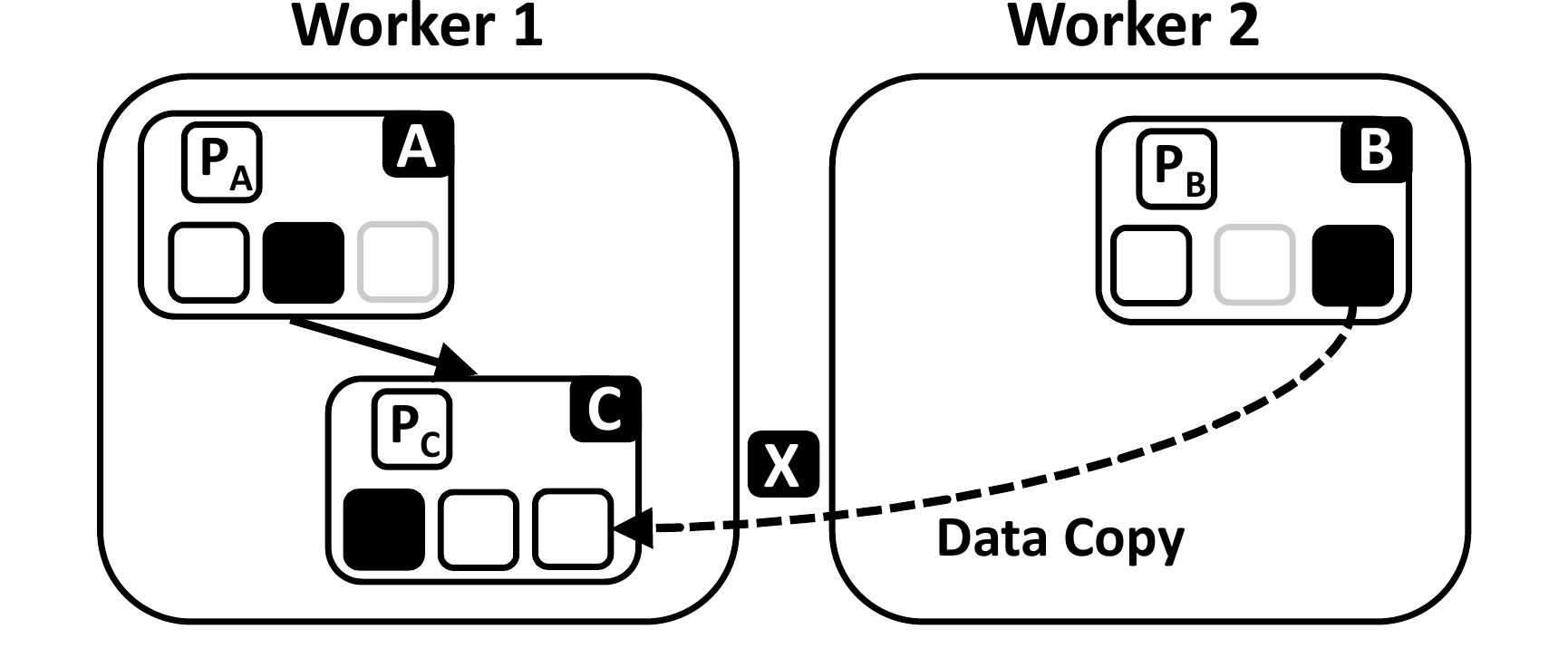}
\label{fig:simple-binding}
}
\end{tabular}
\caption{Simple task graph example (a) and how it maps into per worker task graph
in Nimbus (b).}
\end{figure*}

This section describes the design and implementation of execution
templates in a C++ analytics framework we have implemented, called
Nimbus.  We chose to implement execution templates in a new framework
in order to explore their tradeoffs when not limited by prior design
decisions that might conflict with their goals. There is also discussion
on how execution templates can be introduced into existing frameworks.

\subsection{Nimbus} \label{sec:nimbus}

Because execution templates are tightly entwined with a framework's
data and execution model, we first explain the relevant details of
Nimbus.  The core Nimbus implementation is 15,000 semicolons of C++
code.

\subsubsection{Data Model}
Nimbus has a data flow model similar to DryadLINQ~\cite{dryadlinq},
Naiad~\cite{naiad},
and Spark~\cite{spark}. A job is decomposed into {\it stages}. Each
stage is a computation over a set of input data and produces a set of
output data. Each data set is partitioned into many {\it data objects}
so that stages can be parallelized. Each stage typically executes as
many {\it tasks}, one per object, that operate in parallel. In
addition to the identifiers specifying the data objects it accesses,
each task can be passed parameters, such as a time-step value or
constants.

Unlike Spark's RDDs, and to avoid the cost of data copying noted in
Section~\ref{sec:speedups}, Nimbus allows tasks to mutate data in
place. Mutable data has the additional benefit that multiple
iterations of a loop can access the same objects and reuse their
identifiers. This makes templates more efficient to parameterize, as
the object identifiers can be cached rather than recomputed on each
iteration. There can be multiple copies of a data object. However,
since objects are mutable they are not always consistent. If one worker
writes to its copy of an object, other workers who later read it
will need to receive the latest update.

Data flow between tasks forms a directed acyclic graph (DAG), called
{\it task graph}, whose vertices are tasks and edges are data
dependencies. Figure~\ref{fig:simple-example} shows a simple task
graph with three tasks that operate over three data objects. The
rest of this section uses this example task graph to explain how
templates are generated and instantiated.

\subsubsection{Dependencies and Data Exchange}
\label{sec:explicit}

The goal of execution templates is to allow workers to generate
and correctly schedule large batches of tasks. Not all of these tasks
are immediately runnable. For example, when a worker instantiates the
template in Figure~\ref{fig:simple-example}, it cannot run task C
until both A and B have completed. The ability to locally determine
when it is safe to run a task is critical for reducing load
on a controller; otherwise, the controller would need to publish when
every task has completed. Workers need to be able to know this both when
the dependent tasks are local as well as when they run on another node.

To enforce the correct execution order, each task includes a set of
dependencies. These dependencies are identifiers for tasks that must complete
before the worker schedules the task. As shown in
Figure~\ref{fig:simple-binding}, these dependencies can also be across workers:
task B on worker 2 must complete before task C can run on worker 1. Nimbus
represents this dependency by introducing a pair of control tasks, a send task
on worker 2 and a receive task on worker 1, and inserting the receive task as a
dependency in C. These explicit dependencies allow workers to know when a task
is ready to run without involving the controller.

\begin{figure*}
\centering
\setlength\tabcolsep{-3pt}
\begin{tabular}{cc}
\subfigure[ A controller template represents the common structure of a task
graph metadata. It stores task dependencies and data access patterns. It is
invoked by filling in task identifiers and parameters to each task.]
{
\includegraphics[width=3.0in]{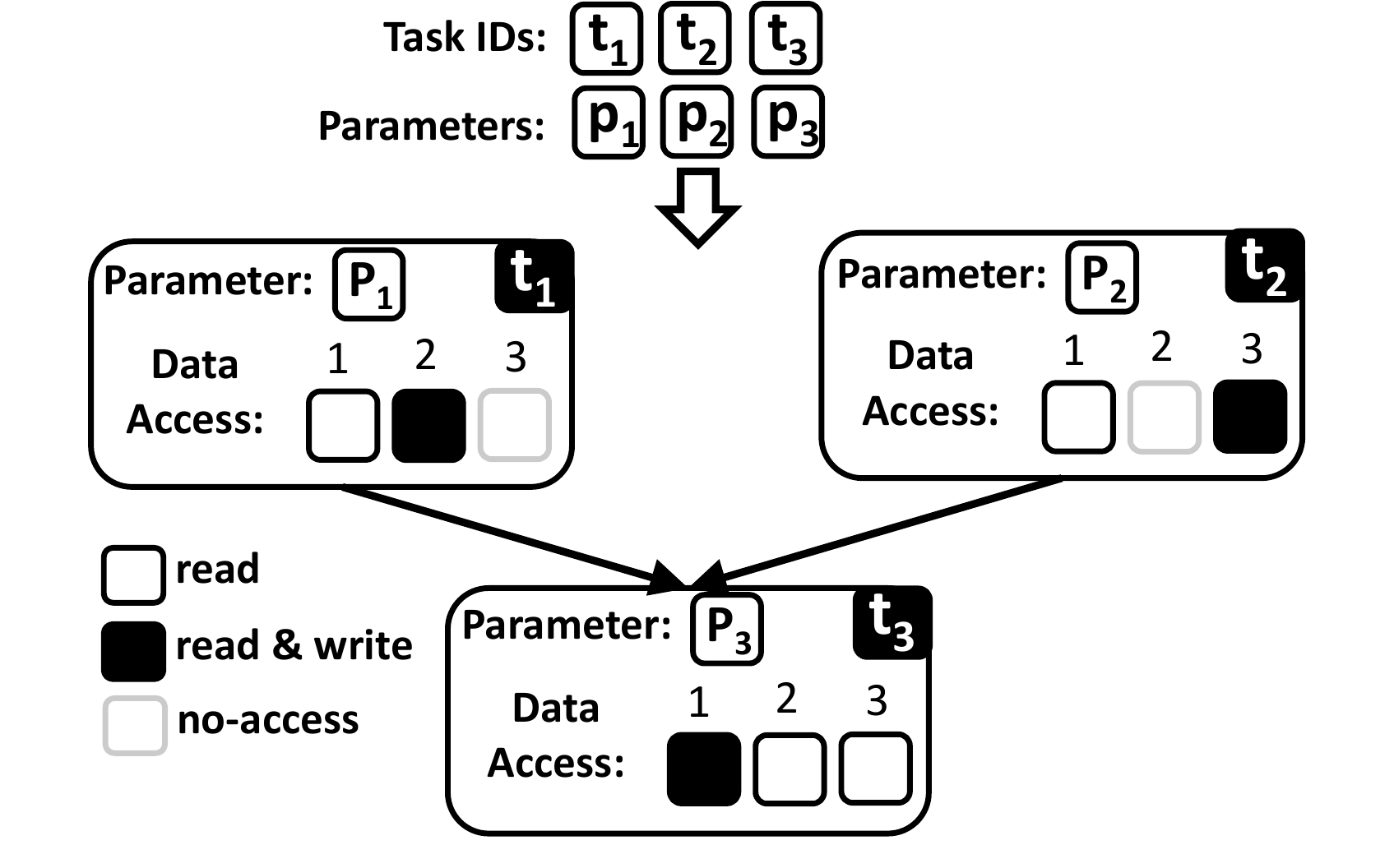}
\label{fig:controller-template}
}
&
\subfigure[Each worker template stores the common structure of a task graph for
execution including the data copies among workers. It is invoked by passing the
task identifiers, and parameters to each task.]
{
\includegraphics[width=3.0in]{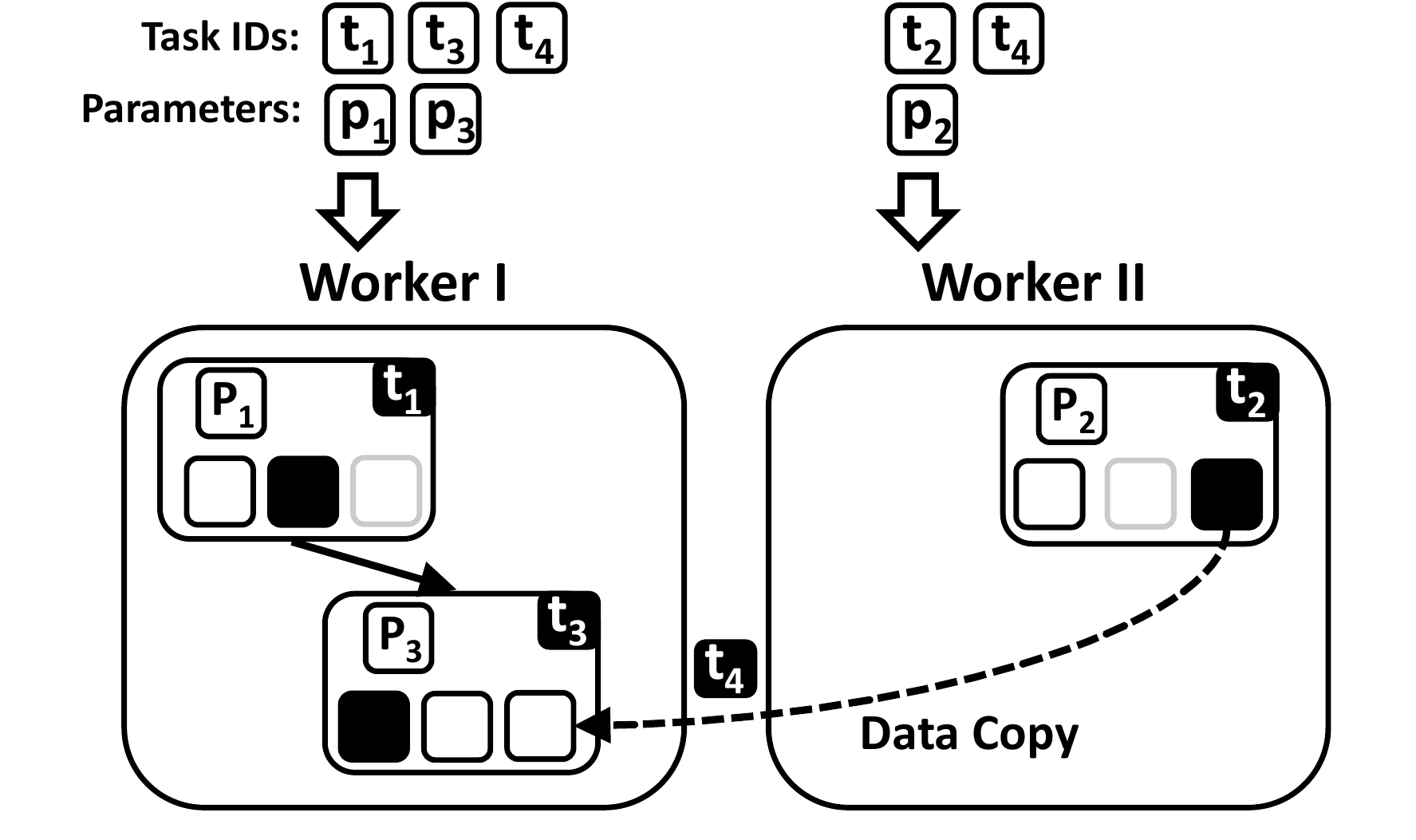}
\label{fig:worker-template}
}
\end{tabular}
\caption{Controller template (a) and worker templates (b) for the task
  graph in Figure~\ref{fig:simple-example}.}
\end{figure*}

\subsection{Dynamic Template Generation} \label{sec:dynamic}

Nimbus generates execution templates on the granularity of basic
blocks.  A {\it basic block} is a code sequence in the driver program
with only one entry point and no branches except at the exit. For
example, in Figure~\ref{fig:driver-code}, there are two basic blocks,
the optimizer and the estimator.  Each iteration of the algorithm
executes the optimizer block (inner loop) multiple times and the
estimator block (outer loop) once.

There are two types of execution templates. {\it Controller templates}
are installed on a controller by the driver program; they encode the
task graph of a basic block across all of the workers. Controller
templates reduce the control overhead between the driver and the
controller. {\it Worker templates} are installed on workers by the
controller; they encode the task graph of a basic block for that
worker.  Once a template is installed, it can be invoked with a single
message. When the driver program invokes a controller template, the
controller typically invokes the corresponding worker template on each
worker.

The two types of templates are decoupled to enable dynamic scheduling
decisions.  If a worker fails or the controller re-balances load across
worker, two invocations of the same controller template can result in
two different partitioning of tasks among workers. For every
partitioning strategy a separate worker template is installed on the
workers.

\subsubsection{Controller Templates}

Controller template stores control dependencies between tasks as
well as which data tasks access.  To create a controller template, the
driver program sends a start message to the controller, which
instructs to start building a template. As the controller receives each subsequent
task, it both schedules it normally and adds it to the
template.  At the end of the basic block, the driver sends a finish
message to the controller. At this point the controller processes the
template it has built and generates worker templates from it. For the
successive instance of the same basic block, driver only invokes the
installed template by passing the task identifiers and parameters to
each task.  Figure~\ref{fig:controller-template} shows how the
controller templates for the graph in Figure~\ref{fig:simple-example}
are installed and invoked.

\subsubsection{Worker Templates}

A worker template has two halves. The first half exists at the
controller and represents the entire computation across all of the
workers.  This centralized half allows the controller to cache how the
template's tasks are distributed across workers and which data objects
the tasks access. The second half is distributed among the workers and
caches the per-worker local task graph with dependencies and data
exchange directives.

As with controller templates, to generate a worker template the controller
sends all tasks to workers explicitly.  Workers execute these tasks
and simultaneously build a template.  The next time the driver program
invokes the controller template, the controller invokes the templates
at each worker by passing new task ids, and parameters.
Figure~\ref{fig:worker-template} shows how the controller templates
for the scheduling strategy in Figure~\ref{fig:simple-binding} are
installed and invoked.

\subsection{Patching Worker Templates} \label{sec:patching}

Templates, when generated, assume that the latest updates to data objects are
distributed in a certain way. For example, in Figure~\ref{fig:simple-binding},
the template on worker 2 assumes that data objects 1 and 3  contain the latest
update. Since templates are installed dynamically, the runtime does not know
the complete control structure of the program. It can be that there are code
paths which do not leave the latest update in every object. Put
in other words, the driver may have issued tasks which invalidate the
template's assumptions.  At the same time, the driver program does not know
where data objects are placed or tasks execute, so cannot correct for these
cases.

Because this problem is subtle, we provide an analogy based on JIT
compilers. JIT generated blocks of native instructions assume that
variables are in particular registers. If the registers do not hold
the correct variables when the block of native instructions is
invoked, then move, store, and load instructions must be added so the
registers do hold the correct variables.\footnote{This is one reason
why JITs often operate on function boundaries, since function calling
conventions specify how variables must be laid out in registers.}

Whenever a template is invoked, the controller needs to first {\it
validate} whether the corresponding worker templates will operate on
the latest updates.  If validation fails, the controller patches the
template by inserting control tasks that copy the latest updates
to the objects. For example, if immediately after the template in
Figure~\ref{fig:worker-template} the same template is invoked, then
controller needs to transfer the latest update of first object
(updated by task $t_3$) to worker 2 to satisfy the preconditions. Only
after patching it is safe to invoke the template again.

Validating and patching must be fast, specially when there are many
workers, data objects, and nodes. For example, the complex graphics
application in Section~\ref{sec:physbam} has almost 
one million data objects. Nimbus uses two optimizations to make validation
and patching fast.

First, for the common case of a template executing twice back to back,
the controller ensures that the input objects to a template hold the
latest updates when the template completes. This is especially
important for when there are small, tight loops: the controller can
bypass both validation and patching. Second, for basic blocks that can
be entered from multiple places in the program (e.g., the block after
an if/else clause), the controller generates a separate template for
each possible control flow.

\subsection{Load Balancing and Fault Tolerance}\label{sec:lb-ft}

Nimbus balances load across workers by periodically collecting
performance statistics at the controller. When the controller
detects that certain workers are busier than others, it redistributes
tasks across the workers, regenerating templates for any workers whose
load has changed.

To recover from worker failures, the Nimbus controller periodically
checkpoints system state. To create a checkpoint, the controller
inserts tasks that commit data objects to durable storage as well as
metadata on where in program execution this checkpoint is. If a worker
fails and the system loses the latest update to an object, the controller
halts all tasks on the workers. It restores the lost objects to their
last checkpoint as well as any other objects which have been modified
since that checkpoint. It then restarts execution, regenerating
any worker templates as needed. If the controller fails, it can restart
and restore the entire system from the checkpoint.

\subsection{Templates in Other Frameworks}\label{sec:others}

Templates are a general abstraction that can be applied to many
frameworks. However, the requirements in Section~\ref{sec:templates}
can be simpler to incorporate in some systems than others. For
example, incorporating execution templates into Spark would require
three significant changes to its data model and execution model,
particularly its lazy evaluation and scheduling. First, it would need
to support mutable data objects. When data is immutable, each
execution of a template is on new data object identifiers. Second, the
Spark controller needs to be able to proactively push updates to each
worker's block manager. Otherwise, every access of a new data object
requires a lookup at the controller. Third, in Spark the controller is
completely responsible for ensuring tasks run in the correct order,
and so tasks sent to workers do not contain any dependency
information. Adding execution templates would require adding this
metadata to tasks as well as worker control logic. While these changes
are all quite tractable, together they involve a significant change
to Spark's core execution model and so we are beginning to discuss
this with its developers.

We are have not yet considered adding templates to Naiad since it is no
longer actively supported (the last code commit was Nov 9, 2014).

\section{Evaluation}\label{sec:evaluation}

\begin{figure*}
\centering
\setlength\tabcolsep{-3pt}
\begin{tabular}{cc}
\subfigure[Logistic regression]
{
\includegraphics[width=3in]{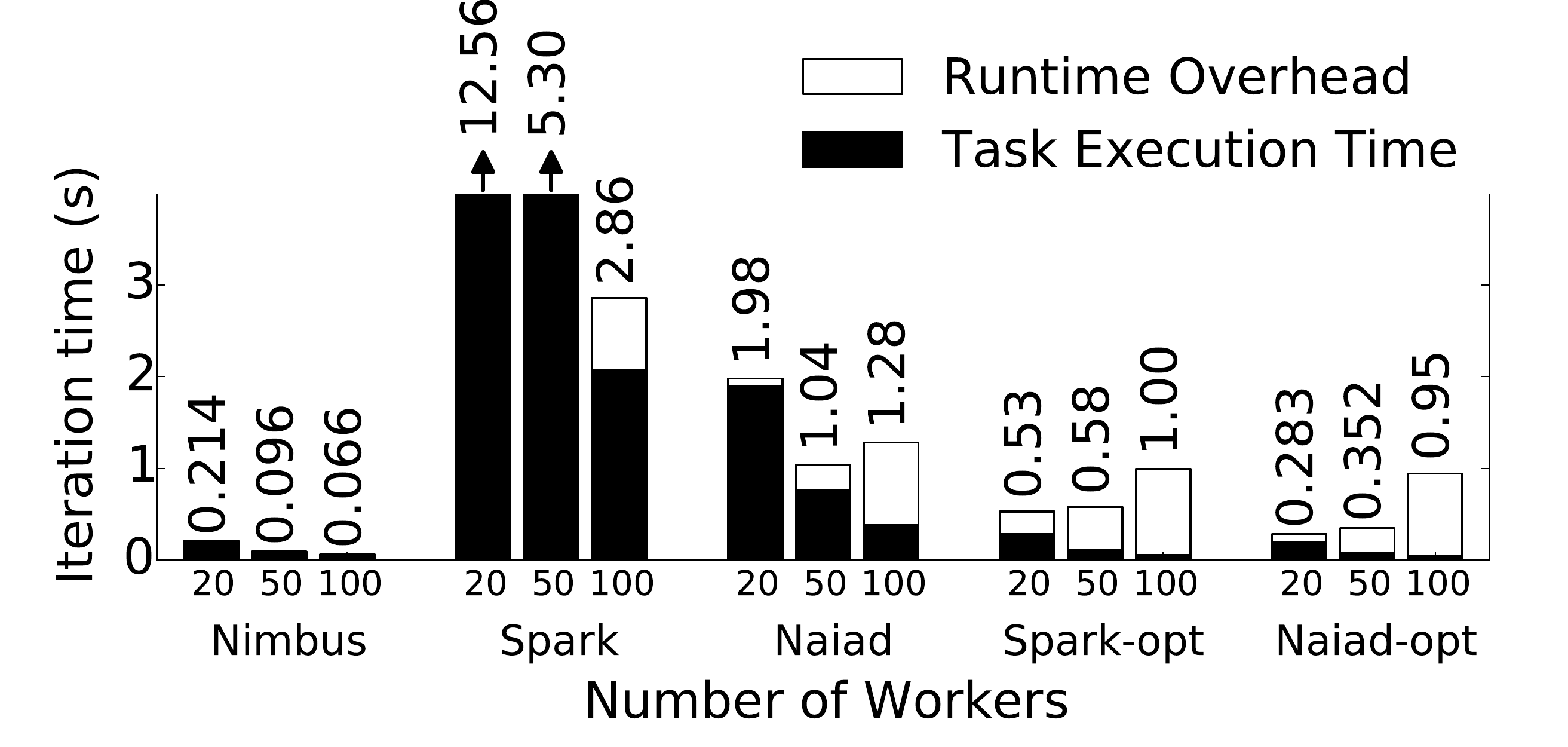}
\label{fig:lr-strong}
}
&
\subfigure[K-means clustering]
{
\includegraphics[width=3in]{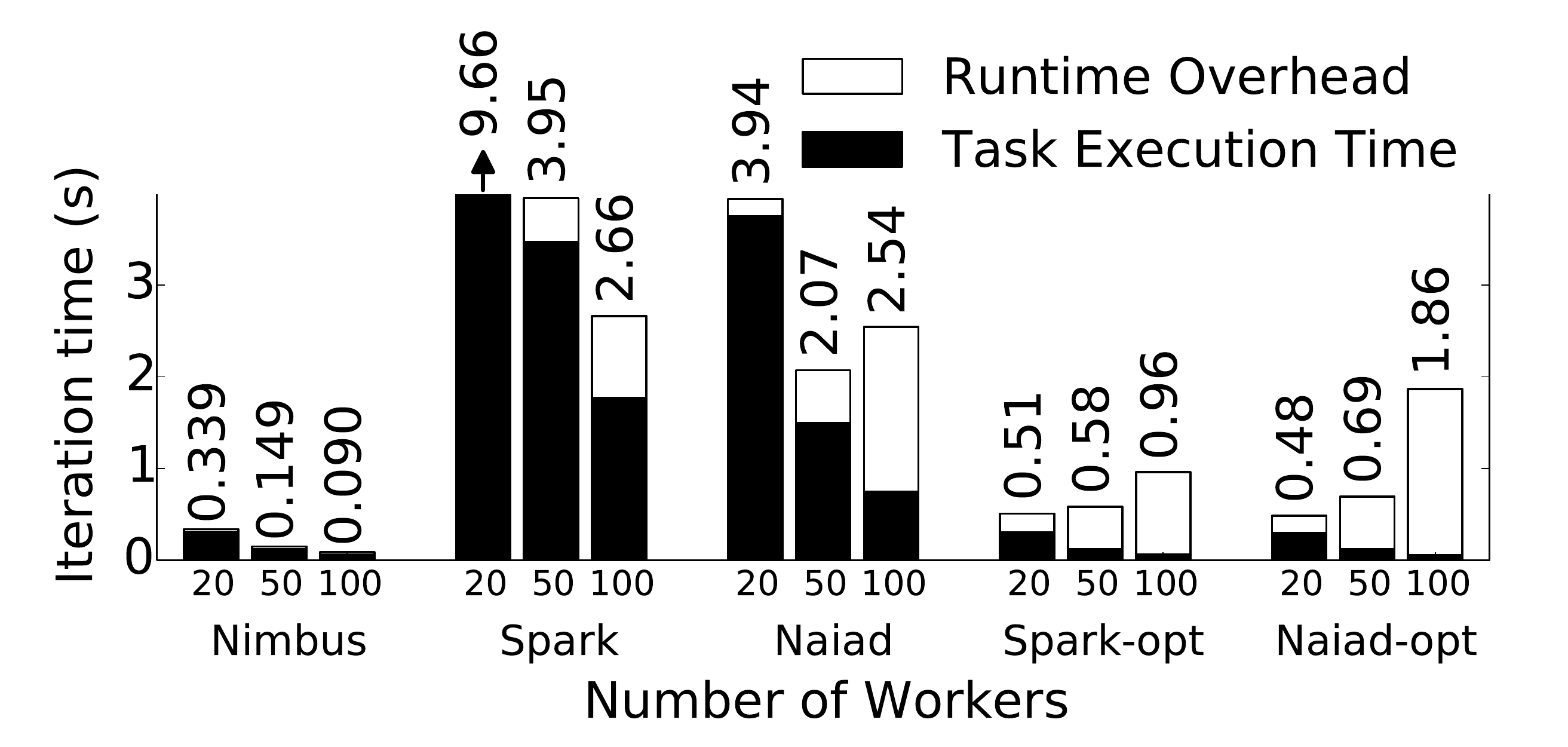}
\label{fig:kmeans-strong}
}
\end{tabular}
\caption{Iteration time of logistic regression and k-means for a data set of
size 100GB.  Spark, Naiad and Nimbus, run Scala, C\# and C++ respectively.
Spark-opt and Naiad-opt show the performance when the computations are replaced
with spin-wait as fast as tasks in C++. Execution templates helps Nimbus scale
out almost linearly.}
\label{fig:data-analytics}
\end{figure*}

This section evaluates how execution templates can support fast,
optimized data analytics jobs at scale. It compares the performance of
k-means and logistic regression benchmarks implemented in Nimbus with
implementations in Spark and Naiad. It measures the costs of computing
and installing templates as well as the performance effect of needing
to recompute worker templates due to load re-balancing. Finally, it
evaluates how far execution templates can scale by measuring their
effect on a distributed graphics workload whose median task length is
13ms and 10th percentile task length is 3ms.

In summary, our findings show:
\begin{itemize}[leftmargin=*] 
  \setlength{\itemsep}{.5pt}
\item Execution templates support orders of magnitude more tasks per
  second than existing centralized (Spark) and decentralized (Naiad)
  designs.  Task throughput scales almost linearly with the number of
  workers.

\item Using execution templates, Nimbus is able to run logistic
  regression and k-means benchmarks 16-43 times faster than Spark and
  Naiad implementations.

\item Half of this performance benefit is from optimized tasks, the
  other half is from execution templates scheduling optimized
  tasks at scale.  If Spark and Naiad use optimized tasks, they cannot
  scale out past 20 nodes; execution templates allow Nimbus to scale
  out to at least 100 nodes and cut completion times by a factor of
  4-8.

\item Using execution templates, Nimbus is able to run a complex
  graphical simulation with tasks as short as $100\mu$s within
  15\% of the performance of a hand-tuned MPI implementation. Without
  templates, completion time increases by 520\% as the
  controller cannot schedule tasks quickly enough.
\end{itemize}

All experiments use Amazon EC2 compute-optimized instances since they
are the most cost effective for compute-bound workloads. Worker nodes
are {\tt c3.2xlarge} instances, which have 8 virtual cores and 15GB of
RAM.  Because we wish to evaluate how the controller can become a
bottleneck, we run it on a more powerful instance than the workers, a
{\tt c3.4xlarge} instances, with 16 cores and 30GB of RAM. This shows
the performance of the controller even when it has more resources than
the workers. We measure completion time of different jobs on 20--100
worker nodes. Nodes are allocated within a placement group and so have
full bisection bandwidth.

Iteration time is averaged over 30 iterations and excludes the first
iteration due to its overhead of data loading and JIT compilation.  We
observed negligible variance in iteration times. For Nimbus, the first
iteration includes the cost of template installation. We therefore
quantify this cost separately from overall performance.

\subsection{Data Analytics Benchmarks}\label{sec:data-analytics}

Figure~\ref{fig:data-analytics} shows the completion time for logistic
regression and k-means when run in Spark, Naiad and Nimbus. In
addition to a Scala implementation in Spark and a C\# implementation
in Naiad, we also measure performance if these frameworks could
execute tasks as quickly as Nimbus. We consider the {\it best case}
performance of no overhead for invoking native code by having them run
a busy loop.

For logistic regression, Naiad's C\# runs 6 times faster than Spark's
Scala. The fastest Spark configuration is 100 nodes, while for Naiad
it is 50 nodes. This is because Naiad's faster task execution means
its control plane overhead overwhelms the benefits of running on more
workers. Naiad's control overhead grows quickly because it requires
O($n^2$) communication among Naiad nodes, where $n$ is the number of nodes.

C++ tasks run 51 times faster than Scala and 9 times faster than C\#.
When Spark and Naiad's tasks are replaced by tasks running as quickly
as C++ code, neither scale out past 20 nodes. We ran them on fewer than 20
nodes: 20 is the fastest configuration. For example, running on 100
nodes, Naiad-opt runs almost 3 times slower than on 50 nodes, as its
$n^2$ coordination overhead grows.

Nimbus runs 43 times faster than Spark and almost 16 times faster than
Naiad. Its control overhead is almost negligible, even when scaled out
to 100 nodes. This allows it to come very close to the expected
performance benefits of C++.  Even if Spark and Naiad were to run
optimized tasks, execution templates lead Nimbus to run 4-8 times
faster.

K-means shows similar results to logistic regression: Nimbus runs
almost 30 times faster than Spark with Scala and 23 times faster than
Naiad with C\#. It runs 5 times faster than Spark or Naiad even when
they use optimized tasks.

\subsection{Task Throughput}\label{sec:throughput}

\begin{figure}
\centering
\includegraphics[width=3.0in]{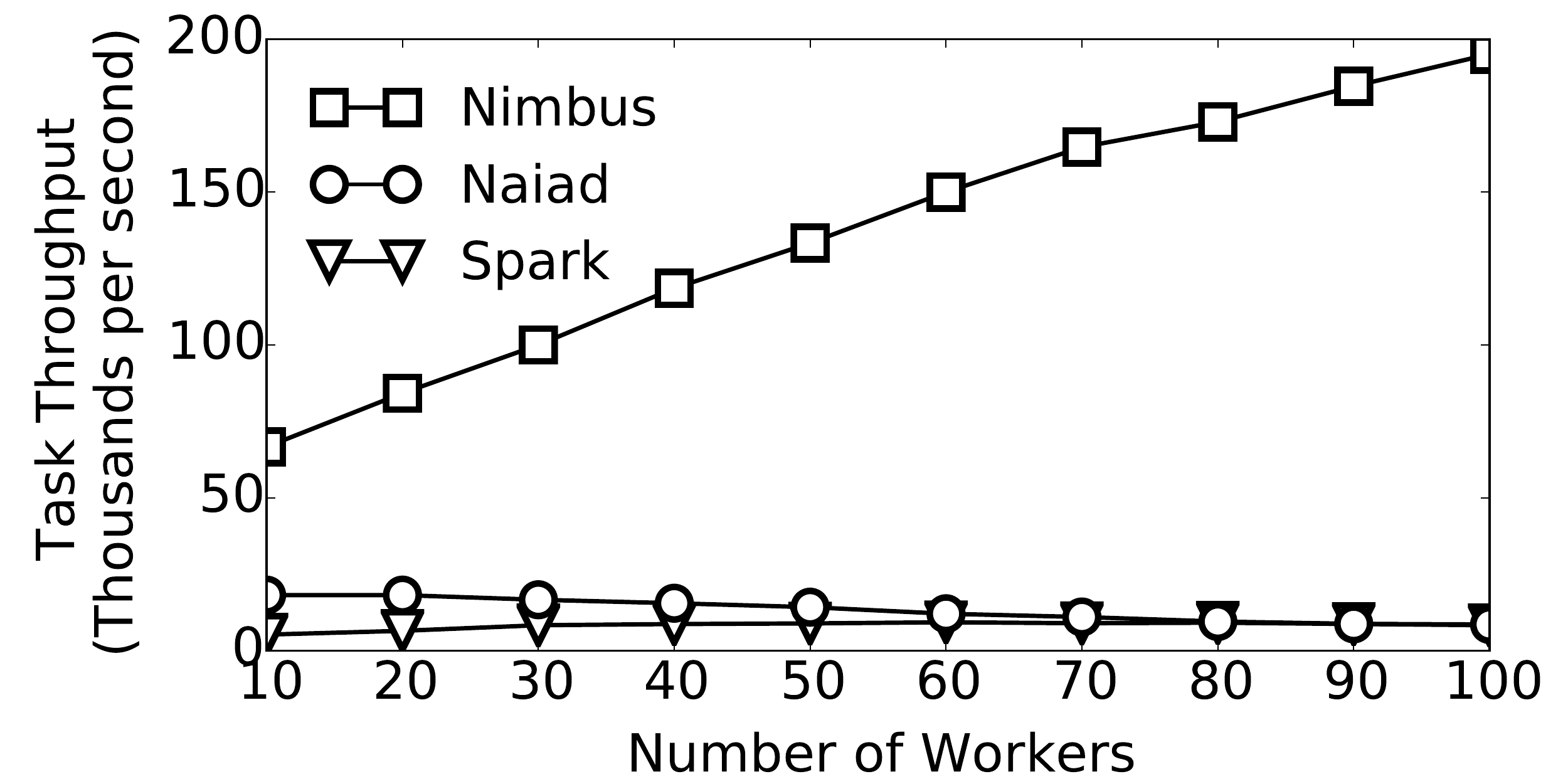}
\caption{Task throughput of cloud frameworks as the number of workers
increases.  Spark and Naiad saturate at about 8,000 tasks per second, while Nimbus
grows almost linearly as the number of workers increases.}
\label{fig:weak-throughput}
\end{figure}

\begin{figure*}
\centering
\includegraphics[width=6.0in]{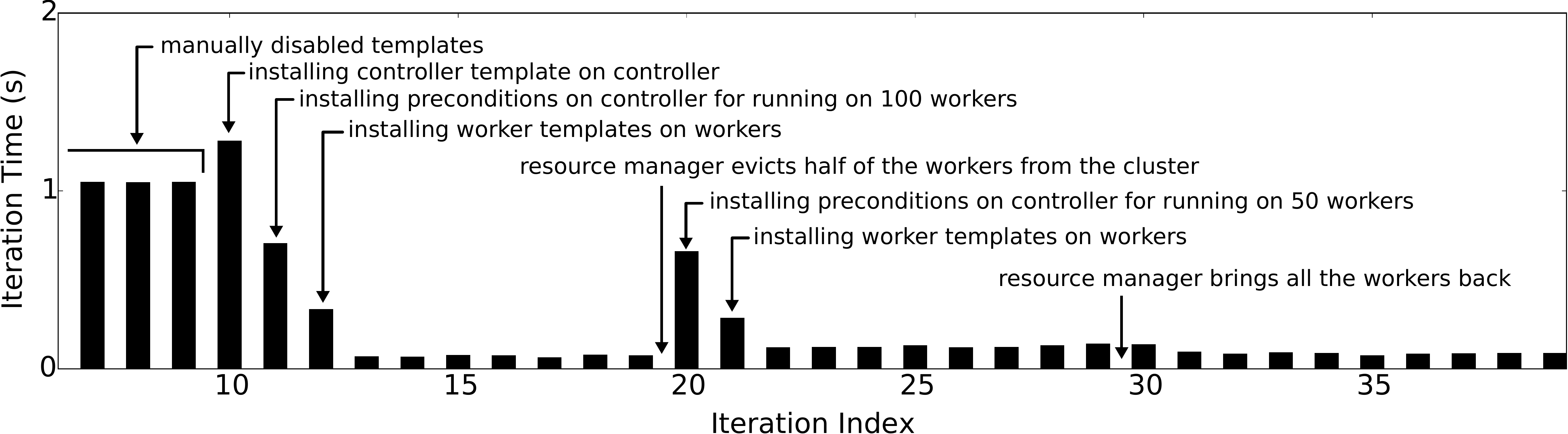}
\caption{Adaptive behavior of execution templates as resources change.
If the number of available workers changes, a controller can recompute
new templates or fall back to templates it has already computed.}
\label{fig:multi-tenant-annotated}
\end{figure*}

The results in Figure~\ref{fig:data-analytics} show that neither Naiad
nor Spark can scale out to handle optimized tasks at scale. Since
progress bottlenecks at the controller, workers spend a
larger fraction of time idle. Figure~\ref{fig:weak-throughput} shows
the {\it task throughput} (the number of tasks per second that workers
execute) each system sustains for logistic regression. Both Spark and
Naiad saturate at about 8,000 tasks per second. Using execution
templates, Nimbus is able to scale almost linearly, supporting almost
200,000 tasks/second for 100 nodes.

Execution templates scale slightly sub-linearly because the scheduling
cost at the controller increases linearly with the number of workers.
If these benchmarks were run on 800 workers with 1 core each (rather
than 100 workers with 8 cores each), each worker template would be
1/8th the size and the controller would have to process 8 times as
many template instantiation messages. If $T$ is the number of tasks to
execute in a block and $W$ is the number of workers, Spark's
controller cost is $O(T)$, Naiad's is $O(W^2)$ and execution templates
are $O(W)$.

\subsection{Template Overhead and Gains}
\label{sec:gain}

\begin{table}
\centering
{\small 
\begin{tabular}{lrr} \toprule[2pt] 
                        & {\bf Per-task cost} & {\bf Iter. overhead}\\ \midrule[1pt]
Controller Template     & $25\mu s$           & 20\%\\
Worker Template @cntrl  & $15\mu s$           & 12\%\\
Worker Template @work   & $9\mu s$            & 7\%\\
\bottomrule[2pt]
\end{tabular}
}
\caption{Costs of installing templates on
  the first iteration of logistic regression running on 100
  nodes. The cost is predominantly at the controller. 
  Nonetheless, the one-time cost of installing templates on
  the first iteration causes the iteration to run 39\% slower.}
\label{tab:costs}
\end{table}

\begin{table}
\centering
{\small
\begin{tabular}{lrr} \toprule[2pt]
                                  & {\bf Completion time} \\ \midrule[1pt]
No templates                      & 1.07s \\
Controller template only          & 0.49s \\
Worker \& controller template     & 0.07s \\ \bottomrule[2pt]
\end{tabular}
}
\caption{Execution time of logistic regression iterations (100 nodes)
  with and without templates.}
\label{tab:cost-gain}
\end{table}

To filter out the startup cost of the JVM and CLR loading object files
and just-in-time compilation, the results in
Figure~\ref{fig:data-analytics} do not include the first iteration of
either computation. This also excludes out the cost of generating and
installing templates. Table~\ref{tab:costs} shows the costs of
installing templates in logistic regression with 100 workers.

Installing templates increases the execution time of the first
iteration by 39\%. This cost is predominantly at the controller, as it
must generate both the controller template as well as the controller
half of the worker template. Processing each task at the controller
takes 40$\mu$s. A controller is therefore limited to processing at
most 25,000 tasks/second on the first iteration: this is approximately
3 times what available controllers can handle.

Table~\ref{tab:cost-gain} shows how controller and worker templates
reduce control plane costs. Both controller and worker templates cut
the overhead significantly as they transform thousands of tasks into a
single message. Their benefits are roughly equal. A controller
template transforms tens of thousands of messages from the driver to
the controller to a single message. Worker templates transform tens of
thousands of messages from the controller to the workers to one
message per worker. Together, they reduce control plane overhead from
93\% to negligible.

\subsection{Template Adaptation}\label{sec:adaptive}
If a controller decides to re-balance a job across workers, remove workers, or
add workers, it must recompute new worker task graphs and install the
corresponding templates on workers whose responsibilities have changed.
Figure~\ref{fig:multi-tenant-annotated} shows the time it takes for each
iteration of logistic regression as a cluster manager adjusts the available
workers. The run begins with templates disabled: iterations take 1.07s. On
iteration 10, templates are turned on. This iteration takes 1.2s due to
controller template installation (20\% overhead). On iteration 11, the
controller's half of worker templates is installed. On iteration 12, the
worker's half of the worker templates is installed. We intentionally separated
each phase of the template installation on progressive iterations to show the
cost and gain from each.  However, all phases could overlap on a single
iteration (39\% overhead).  Once all templates are installed, the iteration
time drops to 0.07s.

On the 20th iteration, the controller receives a command from a cluster manager
to stop using half of its workers. This does not change the controller
template, but forces the controller to recompute worker templates.  It then
executes at half speed (0.14s/iteration), until iteration 30. At iteration 30,
the 50 workers are restored to the controller. It is then able to go back to
using its first set of templates, which are still cached.

\subsection{Complex Applications}
\label{sec:physbam}

\begin{figure}
\centering
\includegraphics[width=3in]{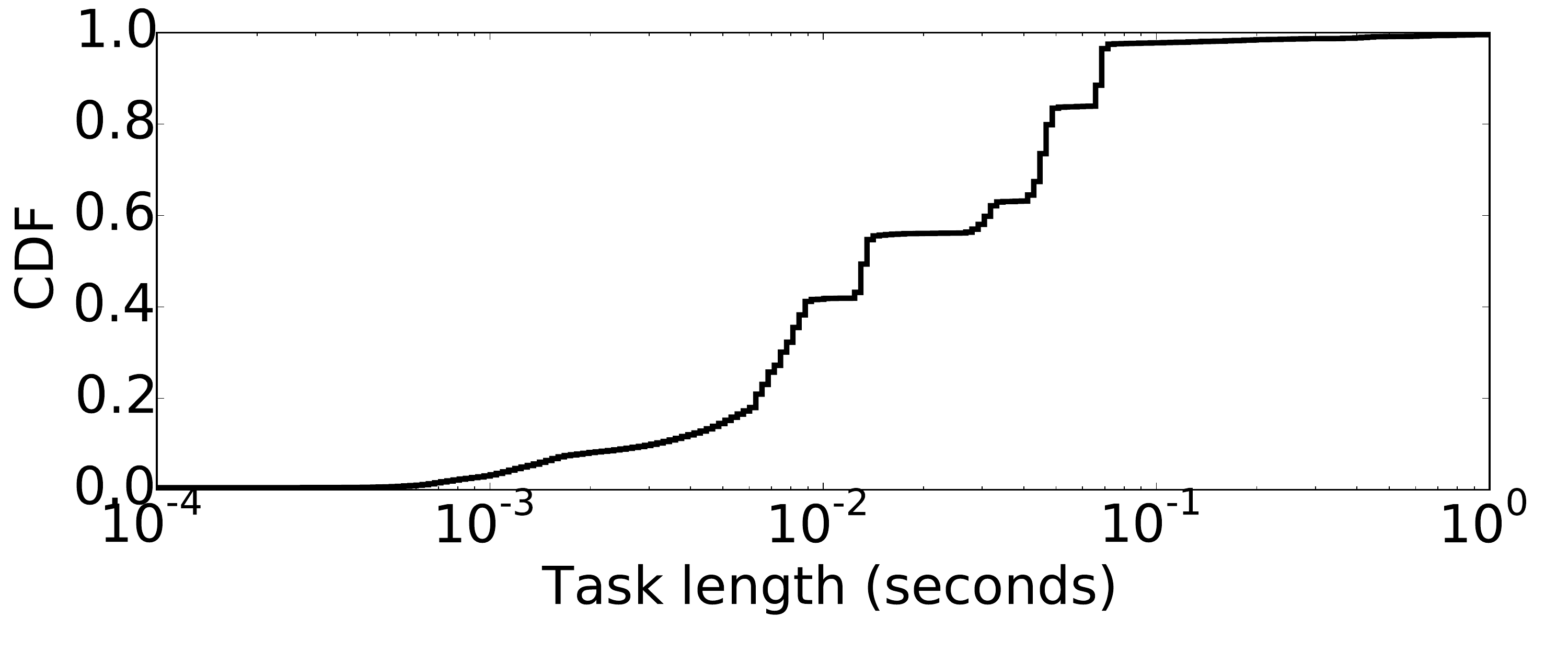}
\caption{CDF of task durations in a PhysBAM simulation. The
 median task is 13ms, the 10th percentile is 3ms, and some tasks are
as short as 100$\mu$s.}
\label{fig:physbam-tasks}
\end{figure}

\begin{figure}
\centering
\includegraphics[width=2in]{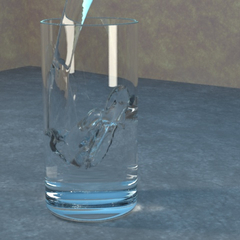}
\caption{Still of a PhysBAM simulation of water being poured into a glass.}
\label{fig:glass}
\end{figure}

\begin{figure}
\centering
\includegraphics[width=3in]{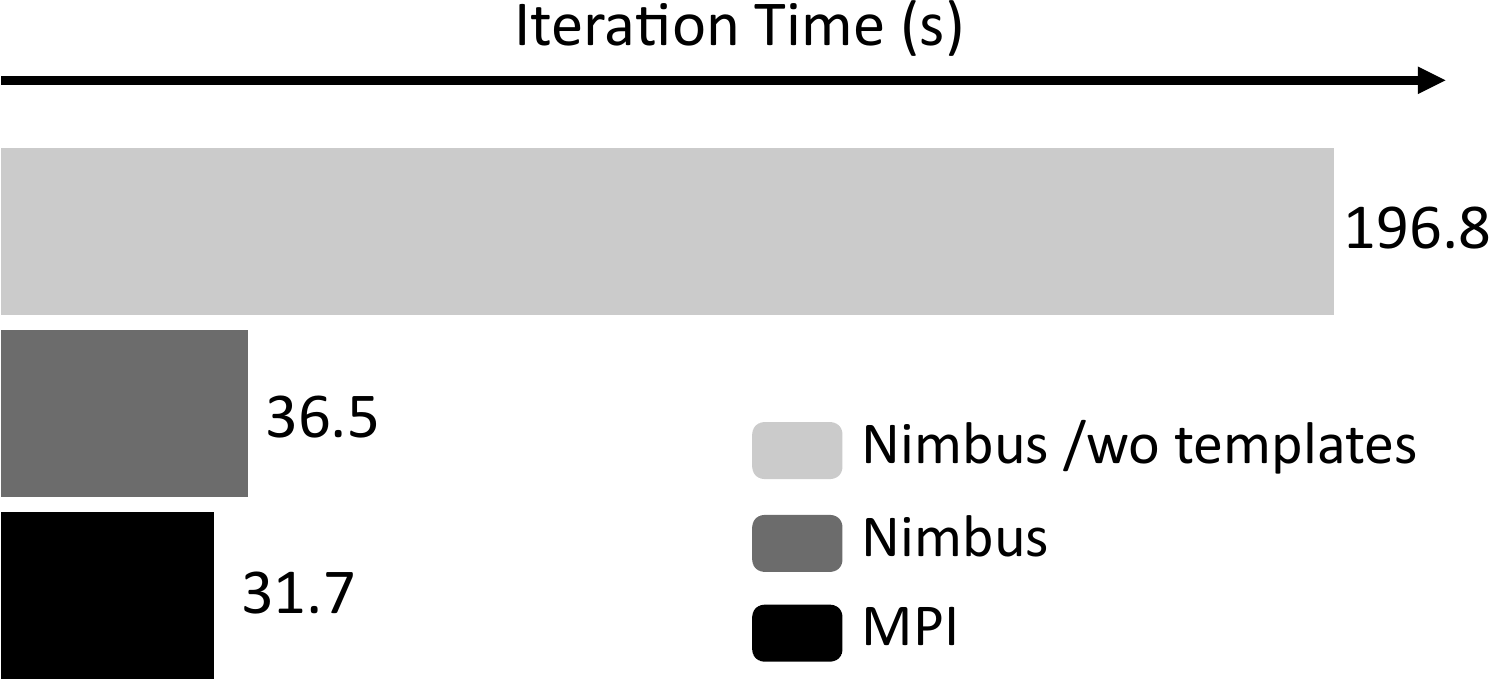}
\caption{Iteration time of a PhysBAM water simulation in Nimbus with
  and without templates as well as its standard MPI implementation.}
\label{fig:physbam}
\end{figure}

This final set of experiments examines how templates scale to support
complex applications. PhysBAM is an open-source library for simulating
many phenomena in computer graphics~\cite{physbam}. It is the result
of over 50 developer-years of work and has won two Academy Awards. We
ported PhysBAM to Nimbus, wrapping PhysBAM functions inside tasks and
interfacing PhysBAM data objects into Nimbus so they can be copied and
transferred.

We wrote a driver program for a canonical fluid simulation benchmark,
water being poured into a vessel (e.g., Figure~\ref{fig:glass}).  This
simulation uses the particle-levelset method~\cite{particle-levelset},
maintaining the simulation as a volume of fixed grid cells but using
particles along the surface of the water to simulate it in much higher
detail. The simulation is the same core simulation used in films
such as The Perfect Storm and Brave and has a triply-nested loop with
26 different computational stages that access over 40 different variables.  

We ran a $1024^{3}$ cell simulation (512GB-1TB of RAM) on 64 workers,
comparing the performance of Nimbus with PhysBAM's hand-tuned MPI
implementation. The MPI implementation cannot re-balance load, and in
practice developers rarely use it due to its brittle behavior and lack
of fault tolerance.

Figure~\ref{fig:physbam-tasks} shows the CDF of task duration in PhysBAM. While
the main computational tasks are 60-70ms some tasks run for only 100$\mu$s.
These tasks computing minimum and maximum values over small sets.
Figure~\ref{fig:physbam} shows PhysBAM's performance using Nimbus and MPI.
Without templates, the simulation generates tasks 8 times faster than a
controller can handle: Nimbus takes 520\% longer than MPI, because controller
becomes a bottleneck. With templates, it runs within 15\% of the MPI
implementation.

\section{Related Work}\label{sec:related}

This paper builds on a long history of related work from several disparate
fields: cloud computing, high performance computing, and programming languages.

\medskip\noindent\textbf{Fast data analytics:}
Within the database and parallel computing communities, prior work has explored the
computational inefficiency of Spark code, proposing new programming models and
frameworks to replace it~\cite{dimwitted,dmll}.  Facebook's AI research group
has open-sourced GPU modules for the Torch machine learning
framework.~\cite{fair}. There is also ongoing research on a common intermediate
language for Spark that provides a glossary of data-parallel optimizations
(including vectorization, branch flattening and, prediction), suggesting
performance in some cases even faster than, hand-written C~\cite{nvl-mit,
nvl-platformlab}.
The trend shows that the next generation of
cloud computing frameworks will execute tasks which run orders of
magnitude faster than today.

\medskip\noindent\textbf{Cloud programming frameworks:}
MapReduce~\cite{mapreduce} is a widely used programming model for processing
large data sets. Open source MapReduce frameworks such as Hadoop and
Hive~\cite{hadoop, hive} are I/O bound: they fetch stable input data from disks, and
save intermediate and final results to disks.  Spark~\cite{spark} uses resilient
distributed datasets (RDDs) to perform computations on in-memory data, while
providing the reliability that data on disk provides. For optimized data analytics with
short task, however, Spark's centralized runtime system becomes a bottleneck.
While Nimbus also uses a centralized controller, execution templates enable
Nimbus to handle orders of magnitude higher task rate.

Naiad~\cite{naiad} is another framework for in-memory computations.
While the distributed
event-based runtime helps scalability without creating a centralized
bottleneck, the cost of synchronization dominates as the number of workers
grows. Logical to physical graph translation on Naiad nodes resembles the worker
templates on Nimbus, however the lack of centralized controller to resolves the
inter-worker dependencies leaves the burden of synchronization on the runtime system.

Dataflow frameworks such as Dryad~\cite{dryad}, DryadLINQ~\cite{dryadlinq},
CIEL~\cite{ciel} and FlumeJava~\cite{flumejava} focus on abstractions for
parallel computations that enable optimizations and high performance.  This
paper examines a different but complementary question: how the runtime scales
out to support very fast computations. In fact, our framework implementation
that incorporates execution templates, Nimbus, resembles the data flow model in
DryadLINQ~\cite{dryadlinq}.

\medskip\noindent\textbf{Distributed scheduling systems:}
There is a huge body of work on distributed scheduling.  They deploy various
mechanisms to provide efficient scheduling decisions with high throughput.  For
example, Sparrow~\cite{sparrow} uses a stateless scheduling  model based on
batch sampling and late binding. Omega~\cite{omega}, on the other hand,
leverages a shared global state through atomic transactions to improve the
allocation decisions. Apollo~\cite{apollo} benefits from a similar model, and
adds task completion time estimations to optimize the scheduling decisions.
Tarcil~\cite{tarcil} is a hybrid model based on both sampling and performance
estimation. Hawk~\cite{hawk}, and Mercury~\cite{mercury} suggest a hybrid
distribute/centralized solution to realize better efficiency in the cluster.

At a very high level, all these systems solve the same problem as execution
templates do: providing higher task throughput at the runtime system. However
there is a very important and subtle difference: these systems distribute the
scheduling across the job boundaries. For a single job with high task rate the
scheduling still goes through a single node. The distributed solution only
solves the problem of multiple jobs producing high aggregate task rate in the
cluster by directing the scheduling of each job to a different node. In a way,
execution templates are orthogonal to theses systems. Every node in the
distributed implementation could benefit from execution templates to support
jobs with orders of magnitude higher task rate.


\medskip\noindent\textbf{High performance computing (HPC):}
MPI~\cite{mpi} provides an interface to exchange messages between parallel
computations, and is the most commonly used framework to write distributed
computations in the HPC domain.  MPI does not include any support for
load-balancing or fault recovery.  Frameworks such as Charm++~\cite{charmpp}
and Legion~\cite{legion} provide abstractions to decouple control flow,
computation and communication, similar to cloud data flow frameworks. Their
fundamental difference, however, is that they provide mechanisms and very
little policy; applications are expected to decide on how their data is placed
as well as were tasks run.  The scale and cost of the machines they are
designed for (supercomputers) is such that they demand more programmer effort
in order to achieve more fine-tuned and optimized use of hardware resources.

\medskip\noindent\textbf{Just-in-time (JIT) compilation:} 
Finally, the idea of memoizing control flow and dynamic decisions in an
execution path closely resembles the approach taken in just-in-time (JIT)
compilers~\cite{dynamo} as well as the Synthesis kernel~\cite{synthesis}.  Both
of these approaches note that particular decisions, while dynamic in the
general case, might lead to deterministic results in any particular case.
Therefore, optimizing that deterministic result can remove all of the
surrounding overhead. While a JIT compiler and the Synthesis kernel generate
optimized native code for particular functions, execution templates generate
optimized structures for scheduling distributed computations.

\section{Conclusion And Future Work}\label{sec:conclusion}

This paper presents execution templates, a novel abstraction for cloud
computing runtime systems that allows them to support extremely high task
rates. The need for high task rates is driven by the observation that many
modern workloads are CPU-bound, and rewriting them in high performance code can
easily lead to task rates that overwhelm modern schedulers. Long-running
applications with high task rates, however, usually consist of many executions
of a loop. Rather than reschedule each iteration of the loop from scratch,
execution templates allow a controller to cache its scheduling decisions and
invoke large, complex sets of tasks on worker nodes with a single message.
Using execution templates, the paper shows that some benchmark applications
reimplemented in C++ can run up to 40 times faster; without templates, their
speedup is limited only  a factor of 5. Finally, execution templates enable
whole new classes of applications to run in the cloud, such as high performance
simulations used in computer graphics.

\bibliography{nimbus}
\bibliographystyle{abbrv} 

\end{document}